\documentclass[12pt]{spieman}  
\usepackage{amsmath,amsfonts,amssymb}
\usepackage{graphicx}
\usepackage{setspace}
\usepackage{tocloft}
\graphicspath{{./}{Figures/}}

\usepackage{changes}

\def\arcsec{$^{\prime\prime}$}
\def\arcmin{$^{\prime}$}
\usepackage{xspace}
\newcommand\ulap[1]{\vbox\@to\z@{{\vss#1}}}%
\newcommand\dlap[1]{\vbox\@to\z@{{#1\vss}}}%
\newcommand\ho{\ifmmode {\rm HI} \else H{\small I} \fi}
\newcommand\hh{\ifmmode {\rm H_2} \else H$_2$\xspace \fi}
\def\no{\ifmmode {N_{\rm HI}} \else $N_{\rm HI}$ \fi}
\def\nt{\ifmmode {N_{\rm H_2}} \else $N_{\rm HI}$ \fi}
\def\msun{\ifmmode {\rm M_{\odot}}\else $\rm M_{\odot}$\fi}
\def\mpc{\ifmmode {\rm M_{\odot} \ pc^{-2}} \else $\rm M_{\odot} \ pc^{-2}$ \fi}
\def\tra{\ifmmode  \text{HI-to-H}_2\else H{\small I}-to-H$_2$ \fi}
\def\aG{\ifmmode {\alpha G}\else $\alpha G$ \fi}
\def\iuv{\ifmmode {I_{\rm UV}}\else $I_{\rm UV}$ \fi}

\def\sg{\ifmmode \sigma_{g} \else $\sigma_{g}$ \fi}
\def\st{\ifmmode \widetilde{\sigma}_g \else $\widetilde{\sigma}_g$ \fi}
\newcommand\hd{\ifmmode \textrm{HI-dust} \else H{\small I}-dust \fi}
\DeclareMathAlphabet{\pazocal}{OMS}{zplm}{m}{n}
\newcommand\ms{\ifmmode \pazocal{M}_s \else $\pazocal{M}_s$ \fi}
\newcommand\Nm{\ifmmode  N_{\rm M}  \else $N_{\rm M}$ \fi}

\newcommand\aj{\ref@jnl{AJ}}
\newcommand\psj{\ref@jnl{PSJ}}
\newcommand\araa{\ref@jnl{ARA\&A}}
\newcommand\apj{\ref@jnl{ApJ}}
\newcommand\apjl{\ref@jnl{ApJL}}     
\newcommand\apjs{\ref@jnl{ApJS}}
\newcommand\ao{\ref@jnl{ApOpt}}
\newcommand\apss{\ref@jnl{Ap\&SS}}
\newcommand\aap{\ref@jnl{A\&A}}
\newcommand\aapr{\ref@jnl{A\&A~Rv}}
\newcommand\aaps{\ref@jnl{A\&AS}}
\newcommand\azh{\ref@jnl{AZh}}
\newcommand\baas{\ref@jnl{BAAS}}
\newcommand\icarus{\ref@jnl{Icarus}}
\newcommand\jaavso{\ref@jnl{JAAVSO}}  
\newcommand\jrasc{\ref@jnl{JRASC}}
\newcommand\memras{\ref@jnl{MmRAS}}
\newcommand\mnras{\ref@jnl{MNRAS}}
\newcommand\pra{\ref@jnl{PhRvA}}
\newcommand\prb{\ref@jnl{PhRvB}}
\newcommand\prc{\ref@jnl{PhRvC}}
\newcommand\prd{\ref@jnl{PhRvD}}
\newcommand\pre{\ref@jnl{PhRvE}}
\newcommand\prl{\ref@jnl{PhRvL}}
\newcommand\pasp{\ref@jnl{PASP}}
\newcommand\pasj{\ref@jnl{PASJ}}
\newcommand\qjras{\ref@jnl{QJRAS}}
\newcommand\skytel{\ref@jnl{S\&T}}
\newcommand\solphys{\ref@jnl{SoPh}}
\newcommand\sovast{\ref@jnl{Soviet~Ast.}}
\newcommand\ssr{\ref@jnl{SSRv}}
\newcommand\zap{\ref@jnl{ZA}}
\newcommand\nat{\ref@jnl{Nature}}
\newcommand\iaucirc{\ref@jnl{IAUC}}
\newcommand\aplett{\ref@jnl{Astrophys.~Lett.}}
\newcommand\apspr{\ref@jnl{Astrophys.~Space~Phys.~Res.}}
\newcommand\bain{\ref@jnl{BAN}}
\newcommand\fcp{\ref@jnl{FCPh}}
\newcommand\gca{\ref@jnl{GeoCoA}}
\newcommand\grl{\ref@jnl{Geophys.~Res.~Lett.}}
\newcommand\jcp{\ref@jnl{JChPh}}
\newcommand\jgr{\ref@jnl{J.~Geophys.~Res.}}
\newcommand\jqsrt{\ref@jnl{JQSRT}}
\newcommand\memsai{\ref@jnl{MmSAI}}
\newcommand\nphysa{\ref@jnl{NuPhA}}
\newcommand\physrep{\ref@jnl{PhR}}
\newcommand\physscr{\ref@jnl{PhyS}}
\newcommand\planss{\ref@jnl{Planet.~Space~Sci.}}
\newcommand\procspie{\ref@jnl{Proc.~SPIE}}
\newcommand\nar{\ref@jnl{NewAR}}

\title{Hyperion: The origin of the stars \\ A far-UV space telescope for high-resolution spectroscopy over wide fields}

\author[a*]{Erika T. Hamden}
\author[b]{David Schiminovich}
\author[c]{Shouleh Nikzad}
\author[c]{Neal J. Turner}
\author[d,e]{Blakesley Burkhart}
\author[f]{Thomas J. Haworth}
\author[g]{Keri Hoadley}
\author[a]{Jinyoung Serena Kim}
\author[h]{Shmuel Bialy}
\author[c]{Geoff Bryden}
\author[a]{Haeun Chung}
\author[i]{Nia Imara}
\author[a]{Rob Kennicutt}
\author[c]{Jorge Pineda}
\author[a]{Shuo Kong}
\author[c]{Yasuhiro Hasegawa}
\author[j]{Ilaria Pascucci}
\author[k]{Benjamin Godard}
\author[l]{Mark Krumholz}
\author[m]{Min-Young Lee}
\author[n]{Daniel Seifried}
\author[o]{Amiel Sternberg}
\author[n]{Stefanie Walch}
\author[c]{Miles Smith}
\author[c]{Stephen C. Unwin}
\author[c]{Elizabeth Luthman}
\author[c]{Alina Kiessling}
\author[c]{James P. McGuire}
\author[c]{Mina Rais-Zadeh}
\author[c]{Michael Hoenk}
\author[c]{Thomas Pavlak}
\author[a]{Carlos Vargas}
\author[a,p]{Daewook Kim}

\affil[a]{Steward Observatory, Department of Astronomy, University of Arizona, 933 N Cherry Ave, Tucson AZ 85719}
\affil[b]{Department of Astronomy, Columbia University, 550 W 120th St, New York, NY 10025, USA}
\affil[c]{Jet Propulsion Laboratory, California Institute of Technology, Pasadena, CA 91109, USA}
\affil[d]{Department of Physics and Astronomy, Rutgers University,  136 Frelinghuysen Rd, Piscataway, NJ 08854, USA}
\affil[e]{Center for Computational Astrophysics, Flatiron Institute, 162 Fifth Avenue, New York, NY 10010, USA}
\affil[f]{Astronomy Unit, School of Physics and Astronomy, Queen Mary University of London, London E1 4NS, UK}
\affil[g]{Department of Physics \& Astronomy, University of Iowa, 203 Van Allen Hall, Iowa City, IA, 52242, USA}
\affil[h]{University of Maryland College Park, College Park, MD, US}
\affil[i]{University of California, Santa Cruz, 1156 High Street, Santa Cruz, CA 95064}
\affil[j]{Lunar and Planetary Laboratory, University of Arizona, 933 N Cherry Ave, Tucson AZ 85719}
\affil[k]{Paris Observatory, 77, Avenue Denfert-Rochereau, 75014 Paris, France}
\affil[l]{Australian National University, Canberra ACT 2600 Australia}
\affil[m]{Korea Astronomical and Space Sciences Institute, 776 Daedeok-daero, Yuseong-gu, Daejeon 34055, Republic of Korea}
\affil[n]{I. Physical Institute, University of Cologne, Zülpicher Str. 77, 50937 Cologne, Germany}
\affil[o]{Tel Aviv University, P.O. Box 39040, Tel Aviv 6997801, Israel}
\affil[p]{Wyant College of Optical Sciences, University of Arizona, Tucson AZ 85721}

\begin{document} 
\maketitle

\begin{abstract}
We present Hyperion, a mission concept recently proposed to the December 2021 NASA Medium Explorer announcement of opportunity.  Hyperion explores the formation and destruction of molecular clouds and planet-forming disks in nearby star-forming regions of the Milky Way.  It does this using long-slit, high-resolution spectroscopy of emission from fluorescing molecular hydrogen, which is a powerful far-ultraviolet (FUV) diagnostic.  Molecular hydrogen (\hh) is the most abundant molecule in the universe and a key ingredient for star and planet formation, but is typically not observed directly because its symmetric atomic structure and lack of a dipole moment mean there are no spectral lines at visible wavelengths and few in the infrared.  Hyperion uses molecular hydrogen’s wealth of FUV emission lines to achieve three science objectives: (1) determining how star formation is related to molecular hydrogen formation and destruction at the boundaries of molecular clouds; (2) determining how quickly and by what process massive star feedback disperses molecular clouds; and (3) determining the mechanism driving the evolution of planet-forming disks around young solar-analog stars.  Hyperion conducts this science using a straightforward, highly-efficient, single-channel instrument design. Hyperion's instrument consists of a 48 cm primary mirror, with an f/5 focal ratio. The spectrometer has two modes, both covering 138.5-161.5 nm bandpasses. A low resolution mode has a spectral resolution of R$\geq$10,000 with a slit length of 65 arcmin, while the high resolution mode has a spectral resolution of R$\geq$50,000 over a slit length of 5 armin. Hyperion occupies a 2 week long, high-earth, Lunar resonance TESS-like orbit, and conducts 2 weeks of planned observations per orbit, with time for downlinks and calibrations. Hyperion was reviewed as Category I, which is the highest rating possible, but was not selected.

\end{abstract}

\keywords{Ultraviolet astronomy, Spectroscopy, Telescopes, Ultraviolet Spectroscopy}

{\noindent \footnotesize\textbf{*}{Erika T. Hamden},  \linkable{hamden@arizona.edu} }

\begin{spacing}{2}   

\section{Introduction}
\label{sect:intro}  
Hyperion is a FUV spectroscopic mission which was recently proposed to the December 2021 NASA Medium Explorer (MIDEX) opportunity. Hyperion seeks to answer three fundamental questions about our galaxy and how stars and planets form within it. First, what is the relationship between the formation and destruction of molecular hydrogen in molecular clouds and the star formation rate of those clouds? Second, how quickly and by which mechanisms do massive stars disperse molecular clouds? Third, what is the driving mechanism for disk dispersal around young, solar-analog stars? Our current understanding of star formation, the impact of feedback on molecular clouds and galaxies, and the structure and lifetimes of protoplanetary disks are all limited by our inability to directly sense and measure the most dominant component of clouds and disks, molecular hydrogen (\hh).

\hh is the most abundant molecule in the universe and is a critical component in the life-cycle of galaxies, stars, and planets. Despite its prevalence, direct observations of \hh are not the standard method of observing molecular clouds or gas rich disks. \hh, a symmetric molecule, has no dipole moment, and no lines accessible in the rest-frame visible. It does not radiate efficiently when cold because of the lack of dipole. \hh does have significant sets of lines that can be excited via fluorescence. These lines are visible in both the infrared (IR) and UV. IR ro-vibrational lines are typically weak, making them challenging (though not impossible) to observe from the ground. UV-pumped \hh , which is the primary observational tracer for Hyperion's mission, have fluxes directly proportional to the incident exciting UV flux ($F(H_2)_{UV} \sim B_{mn} F(\lambda_{UV})$, where $B_{mn}$ is the branching ratio of the \hh UV transition from the excited electronic band energy $m$ and its decayed energy level in the ground electronic band, $n$), whereas subsequent IR transitions will branch out further, leading to  additional losses in line flux over the IR for a given UV emission line ($F(H_2)_{IR} \sim zF(\hh)_{UV}$\cite{1988Sternberg}, where $z$ is a constant ($\lesssim$0.35) and in the absence of additional contributions to the IR emission line, like collisional effects).

In addition, IR ro-vibrational lines often require special conditions to arise, either in the way that \hh is heated to represent higher thermal states (e.g., shocks) or how hh emits IR photons (e.g., collisional decay). In most situations where IR-\hh lines are observed, the role of UV-pumping is just one of several possible pumping mechanisms, which makes interpretation 
of the initial conditions which initiate the \hh fluorescence process challenging\cite{2018Youngblood, 2021Kaplan}. Indeed, Ref.~\citenum{2021Kaplan} shows that, even in star-forming regions where IR-\hh level populations show the least deviation from pure UV excitation, they always measure a notable deviation from a purely UV excitation population of \hh (most likely due to collisions).

Combined, IR ro-vibration emission requires special boundary and excitation conditions, whereas UV emission probes nominal conditions along PDR and molecular cloud boundaries subject to UV irradiation from nearby, newly formed (OB) stars. A great example comes from the recent image of NGC 3324 from \emph{JWST}. Ref.~\citenum{2022Reiter} dive into the NIRCAM observations of this famous star-forming region, revealing structures illuminated by IR \hh ro-vibrational emission. Of note, however, is that IR \hh emission is observed exclusively in outflows and shocks littering the region. The PDR itself, and of note the molecular cloud bound exhibiting streaming gas from the molecular cloud boundary in the form of hydrogen Pa-$\alpha$ emission, is devoid of IR \hh emission. This type of boundary represents a typical cloud boundary where Hyperion is uniquely able to detect coincident UV \hh features, alongside the other evaporating gas at the molecular cloud boundary, that IR \hh does not appear.

We note that, while the advent of JWST now offers exquisite sensitivity and provides the first near- to mid-IR observatory with the power to uncover IR \hh in ways previously unobtainable, the spectral resolution of JWST's instrument suite makes it inadequate to perform the necessary observations to derive critical physical conditions about the gas that make up Hyperion's suite of science goals. Previous PDR studies using IR \hh have required high spectral resolution instruments\cite{2021Kaplan}, just as Hyperion's UV \hh measurements require high spectral resolution.

The UV lines span both the extreme UV (EUV) and the FUV, but have been observed by a range of space telescopes for limited sightlines. For large area surveys, astronomers have typically used other, less abundant, molecular tracers that are easily accessible and bright in the near IR and millimeter (NIR; i.e. CO, HCN, dust). CO and other tracer emission is then converted to \hh mass using a scale factor (e.g., X$_{CO}$, \cite{2013Bolatto,2013Sandstrom}). These conversion factors can vary but are not necessarily universal on all scales, yielding large uncertainties in mass \cite{2014Lee}. 

Astronomers have known about the importance of \hh for nearly 100 years. \hh in the inter-stellar medium (ISM) was first postulated in 1934 \cite{1934Eddington}, but not detected \cite{1968Werner} until the first UV sounding rockets in the late 1960s \cite{1967Carruthers,1970Carruthers} which yielded the first absorption-line measurements in dense interstellar regions \cite{1971Hollenbach}. Since then, many other observations have confirmed the presence of \hh along sightlines throughout the galaxy, both in absorption \cite{1973Rogerson,1975Spitzer,1976Spitzer,1974Spitzer,1973Spitzer} and emission. Models predicting UV emission from fluorescing \hh \cite{1967Stecher,1962Osterbrock,1963Gould} were first tested in the laboratory \cite{1970Dalgarno} before being observed from an astrophysical object. \hh was then detected in emission towards the reflection nebulae IC63 by IUE \cite{1989Witt} and in the diffuse ISM with the Berkeley UVX Shuttle Spectrometer \cite{1990Martin}, followed by subsequent Shuttle- and rocket-based studies. Additional studies with FUSE \cite{2006Lupu} and others \cite{2005Murthy,2005France,2004France} have detected \hh in a range of targets. Most recently, FIMS/SPEAR revealed galaxy-wide \hh fluorescence \cite{2017Jo}. While these data provide confidence that \hh fluorescence is prevalent throughout most nearby molecular clouds, the existing data sets are not sufficient in spectral or spatial resolution nor in area coverage to complete Hyperion’s objectives.

One of the technical challenges of observing \hh in the UV is that most UV missions have instruments which are count rate limited, and have a maximum brightness cutoff above which the detector will saturate. The effect of this limit is that observations of regions nearby to the UV bright stars that stimulate the brightest fluorescence are not possible. Hyperion is well-suited to covering large areas, for which the observing time metric is inversely proportional to the product of the efficiency, aperture area, and the slit area. By comparison, the only UV spectrographs operating today (HST/STIS and HST/COS), which were optimized for observing point sources and individual sightlines, show that Hyperion has a comparable efficiency, smaller primary aperture area, but significantly larger slit. This means that Hyperion's planned large area survey of the brightest star forming regions is not possible with existing telescopes. A recently flight-tested, high-performance UV solid state detector technology, delta doping \cite{2017Nikzad}, applied to CCDs with low noise and large full well capacity creates a high efficiency, high dynamic range solution that eliminates the problem of count rate limits of bright objects. Future UV mission such as Hyperion can now observe the brightest parts of nearby star forming regions to capture \hh fluorescence, benefiting from the large detector dynamic range.

Hyperion conducts its observations with a simple single channel instrument: a high-resolution FUV spectrograph. Hyperion has a 48-cm two-mirror Cassegrain telescope, with two additional optics to provide field correction at the edges of the field. Hyperion has a single, very long slit (70 arcmin), with a stepped geometry configuration. The slit center is 2 arcseconds wide, giving a resolution R$>$50,000 over 5 arcmin, while the rest of the slit is 10 arcseconds wide, with a resolution of R$>$10,000 over a 65 arcmin field. The current best estimate (CBE) of angular resolution along the slit is $<$4 arcseconds. The spectrograph is an Offner relay design, and includes a folding mirror just before the detector to reduce the impact of radiation on the detector. 

Hyperion’s overall percentage throughput is similar to HST STIS, but Hyperion’s large field-of-view lets it cover hundreds of square degrees of molecular clouds in an 18 month survey mission. Achieving similar coverage with HST would take $>$40,000 years (STIS high-resolution aperture: 0.034\arcsec\ × 28\arcsec\ = 0.952 sq\arcsec\ vs.\ Hyperion narrow aperture: 2\arcsec\ × 300\arcsec\ = 600 sq\arcsec; Hyperion wide aperture: 10\arcsec\ × 3900\arcsec = 39000 sq\arcsec). 

Moreover, the sensitivity of Hyperion itself is two orders of magnitude higher than HST STIS. The emission line sensitivity of Hyperion with wide aperture over a 10\arcsec $\times$ 300 \arcsec area in 2,600 sec is 1$\times$10$^{-17}$ erg s$^{-1}$ cm$^{-2}$ arcsec$^{-2}$ (MEV). In contrast, the emission line sensitivity of HST STIS over 0.034\arcsec\ × 28\arcsec\ area in 2,600 sec is 1.77$\times$10$^{-15}$ erg s$^{-1}$ cm$^{-2}$ arcsec$^{-2}$. These differences are primarily due to the small aperture of HST STIS.
In addition, HST has strict count rate limits that eliminate observations of the UV bright stars and dynamic molecular regions that Hyperion can observe using the high dynamic range detector.

The Hyperion mission was proposed in 2019 as a Small Explorer (SMEX; Choi et al., 2021 \citenum{2021Choi}). The current version of Hyperion was re-formulated for the 2021 Medium Explorer call. The changes include an updated optical design, using only a single grating instead of an additional cross disperser, a larger FPA comprising a 4 detector linear mosaic vs. a single detector, a slightly larger aperture (40 vs. 48 cm), and a different orbit (TESS-like for MIDEX, LEO for SMEX). The MIDEX selections occurred in mid-August, 2022 and while Hyperion was reviewed as a Category I mission, the highest rating possible, it was not selected.

In this paper, we provide a brief overview of the physics of \hh fluorescence (Section \ref{sec:htwo}) and how it can be used to measure physical rates and incident radiation, describe Hyperion's Science Objectives in detail (SOs, Section \ref{sec:Science}), and the Hyperion instrument implementation (Section \ref{sec:instrument}). Section \ref{sec:survey} describes the Hyperion survey strategy over its 18 month mission while Section \ref{sec:other} describes how the Hyperion instrument can be used for science beyond its specific objectives.  

\subsection{Critical science questions that can only be answered in the UV}

Hyperion addresses several foundational science questions that currently can't be answered with existing data. Hyperion is specifically designed to fill both the holes in our current knowledge of molecular clouds and disks and address a missing observational capability. 

Hyperion will determine how star formation relates to the flow of gas into and out of molecular clouds; how massive stars disperse such clouds to end star formation; and how the gas is removed from the planet-forming disks around young stars like our Sun. Hyperion does this by focusing on the atomic-to-molecular transition, which acts as a boundary layer or envelope for a molecular cloud or planet forming disk. This transition layer is the interface between the nursery-like environment within the \hh region and the harsh, radiation-filled outside of a cloud or disk. By observing this layer, we determine the evolutionary states of star-forming molecular clouds, the interior of bubbles blown within molecular clouds by young stars, and in the winds of gas rich planet-forming disks.

Hyperion first addresses the life cycle of \hh molecules and contrasts that with the life cycle of molecular clouds. Is the formation of \hh a critical and necessary step in the path to star formation, or is it just a side effect of the actual evolutionary path? This is discussed in Sec. \ref{sec:so1}. Hyperion also explores the end of a molecular clouds life, as it is dispersed via various feedback mechanisms, in Sec. \ref{sec:feedback}. Finally, Hyperion explores the molecular content of planet forming disks in Sec. \ref{sec:disks}.

\subsection{The diagnostic power of \hh fluorescence}

The key innovation to answering Hyperion's science questions is to use the FUV fluorescence of \hh, which provides diagnostic details on the state of the gas, formation and destruction rates, energy source, and other key pieces of information.

\hh consists of two hydrogen atoms held together by a covalent bond. In the ISM, \hh formation occurs on dust grains, a process first proposed by Ref. \citenum{1948Vandehulst} and calculated by Ref. \citenum{1960McCrea}. Formation occurs from catalytic reactions on the dust surface, and dust and \hh are commonly found together in the ISM. \hh itself is non-polar, a homonuclear and symmetric molecule with no dipole moment \cite{2017Wakelam}. The lack of a permanent diople means that ro-vibrational transitions are forbidden and only weak electric-quadrupole transitions are allowed \cite{1998Wolniewicz}. This causes the ``unfavorable location of its spectrum", from the perspective of Ref. \citenum{1966Field} and many astronomers since. Despite this negative attitude, \hh in fact has a rich spectrum of emission lines in the UV and IR that are excited by fluorescence, typically from an external UV source, exciting the Lyman and Werner bands of \hh. Some emission in the IR can also be excited by collision excitation, which makes analysis of IR \hh emission fairly complex. This confusion is not present in the UV because these collisions don't have enough energy to stimulate the UV bands.

\subsubsection{{\bf Physics of H$_2$ photodissociation and \hh fluorescence}}
\label{sec:htwo}

\begin{figure*}
    \centering
    \includegraphics[width=0.5\columnwidth]{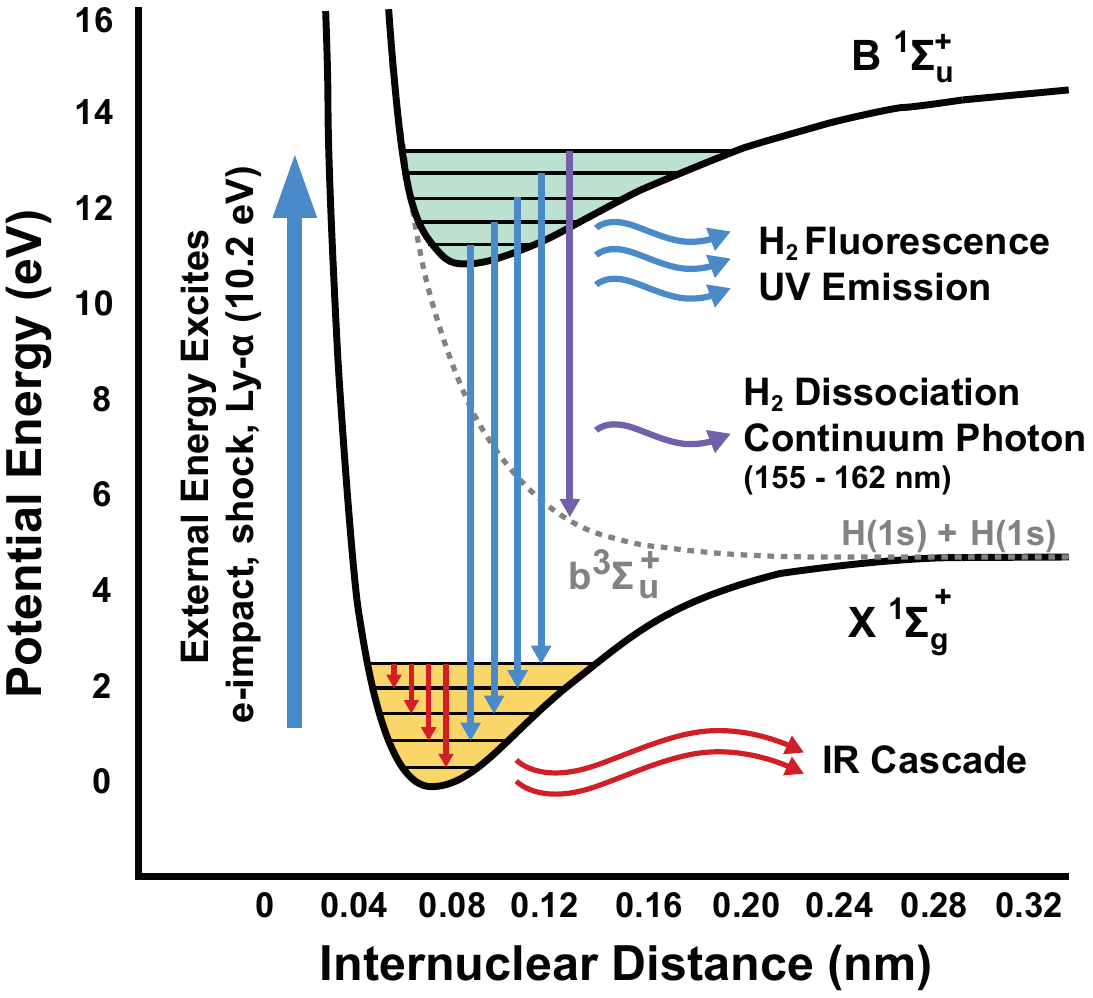}
    \caption{Energy diagram for \hh molecule, with an example excited band}
    \label{fig:h2physics}
\end{figure*}

Fluorescence occurs when an \hh molecule is excited from its ground electronic band (X$^1\Sigma_g^+$) into an excited band (for example, Lyman, the first excited band, B$^1\Sigma_u^+$, up arrow in Figure \ref{fig:h2physics}). When this happens, the molecule will decay back down to the ground level via one of many pathways (blue down arrows), producing a dense line spectrum whose exact wavelengths depend on which sub- states within the electronic bands are involved; each de-excitation carries a ~15\% chance of dissociating the \hh molecule entirely (dissociation continuum, purple down arrow). Subsequent cascades in the ground band release IR photons (red arrows). Excitation into the Werner band, (C$^1\Pi_{\mu}$) creates a different set of emission lines following the same process \cite{1989Sternberg,1989Sternbergb}. 

The exact lines that are excited by \hh fluorescence in the ultraviolet depend strongly on the energy source \cite{1989Sternberg}. This means \hh fluorescence also provides information on the excitation source itself and can be distinguished by different excited progressions. In addition, the overall level of fluorescent emission is related to the incident FUV radiation field at the \hh surface. 
In the Solar neighborhood, the FUV  (integrated over 6-13.6 eV) interstellar radiation field has a flux of $F_0 = 2.7 \times 10^{-3}$ erg cm$^{-2}$ s$^{-1}$ \cite{Draine1978, Draine2011, Bialy2020}.
As we will be exploring clouds of various properties, we use the 
normalized FUV intensity, $I_{\rm UV} \equiv F/F_0$\footnote{In these units, Ref. \citenum{1968Habing}'s estimate of the interstellar radiation field is equivalent to 0.6, and Ref. \citenum{Draine1978}'s field has $I_{\rm UV}=1$ to characterize the field intensity that illuminates the cloud.}

Because both processes of the FUV line emission and the H$_2$ photodissociation are initiated by the same process, namely, the H$_2$ photo-pumping, the H$_2$ photodissociation rate and the summed intensity of all H$_2$ emission lines are proportional to each other (e.g., see Eq. 10 in Ref. \citenum{1988Sternberg}; see also Ref. \citenum{Bialy2020B} for an analogous derivation for the case of cosmic-ray excitation). 
However, because the emission lines are affected by dust extinction, the proportionality between the total line emission power and the photodissociation rate is only exact in optically thin regions, whereas in optically thick gas, one must correct for extinction by dust.
Thus, given an estimate of the dust optical depth, a measurement of the H$_2$ FUV spectrum may be used to estimate the H$_2$ photodissociation rate. An elaborated discussion and numerical tests are presented in Bialy et al. 2022 (in prep.), and in Burkhart et al. 2022 (in prep.).

\subsubsection{Formation rates via ortho-to-para lines}
A key signature found in the \hh UV-fluorescence emission lines is the ortho-to-para ratio (OPR). In the ground state, ortho molecules have spins in the same direction (total nuclear spin=1, only odd rotational quantum numbers j), while para molecules have spins in opposite directions (zero total nuclear spin, only even j states). These different states have subtly different allowed transitions for fluorescence. 

The imprint of \hh formation in the cloud, and how far out of equilibrium \hh formation may have occurred, is traced by the ratio of the \hh fluorescent ortho-para lines, which can be used to infer the OPR in the illuminated gas. The OPR measurement works as a chemical clock because \hh forms on dust grains with an OPR near 3, but this value is far out of thermal equilibrium. Over millions of years, slow collisional processes lead this elevated OPR to decay down to the thermal equilibrium value of ~0.001 for low temperature gas \cite{2006Flower,2013Pagani}; measuring the OPR therefore reveals how long ago a particular population of molecules formed. The exact decay rate depends on ionization, which is well understood for cosmic rays \cite{1984Flower} and modeled for other sources.

For equilibrium conditions in a photodissociation region (PDR), the equilibrium OPR in the ground vibrational state is set by the competition between the reactive collisions with protons, the \hh formation process, and selective photodissociation via optically thick (“self-shielded”) absorption lines (\citenum{1999Sternberg}, see Eqs. 4-7). A low OPR provides no ambiguity in the state of the gas; measuring a low OPR is a sure sign of older, colder \hh gas. A high OPR can be the result of two different processes. One is a warm PDR that has already reached chemical equilibrium, while the other is cold, young \hh gas that has not yet reached chemical equilibrium and has a high OPR due to the formation process on dust grains. A way to distinguish between the two is via a temperature measurement, which can be self-consistently estimated using the H2 fluorescent emission. The temperature (in and out of equilibrium) will be modeled using the Meudon PDR code \cite{2006lepetit}. Hyperion will incidentally collect absorption line measurements in the process of mapping target clouds. This can provide measurements of cold H2 through parts of the cloud, for which the chemical equilibration time is unambiguously long and therefore the OPR measurement can be straightforwardly interpreted as chemical age.

\section{Science Motivation of the Hyperion Mission Concept}\label{sec:Science}

\subsection{Formation: Understanding molecular cloud formation and evolution}\label{sec:so1}


Most interstellar gas is in the form of atomic hydrogen (HI) however ultimately it is the molecular gas, predominately in the form of \hh, that fuels the formation of stars. For example, the seminal results of Ref. \citenum{Bigiel2008AJ....136.2846B} have shown that HI is largely not observed to
correlate with star formation and the strong correlation of the star formation rate surface density  with gas surface density only begins as column densities greater than the atomic to molecular transition \cite{1988Sternberg, Krumholz2009THESF, 2010ApJ...723.1019H,2012ApJ...745...69K, Sternberg2014, Bialy2015, Bialy2016, Bialy2017}. This may imply that the transition between the atomic to molecular phase is important for the formation of stars or it may simply be that the formation of molecular gas a natural step on the pathway for gas to become dense and cold, a pathway which is also necessary for star formation \cite{2012ApJ...759....9K,2016ApJ...829..102I,2018ApJ...859..162C,2019ApJ...881..160B,2019ApJ...879..129B}.  Ultimately, understanding how the lifecycle of molecular gas connects to star formation necessitates observations that capture the HI to \hh transition process in action  \cite{2022arXiv220309570C}.  

With present telescopes and coverage of the electromagnetic spectrum, studying the HI-\hh boundary layer can only be done on individual sight-lines or indirectly using tracers other than \hh itself, e.g. CO \cite{2014PhR...539...49K} and dust \cite{2012Lee}. 
Hyperion provides
exactly this measurement, using the unique
diagnostic of \hh fluorescence which, in an analogous
manner to H$\alpha$ emission in photoionized H$_{II}$
regions, provides a direct tracer of the process of
\hh dissociation at the atomic to molecular boundary and phase transition mass flux which determines if the  boundary is advancing
or receding.


Tracking the formation and destruction
of \hh allows us to determine both how
clouds are assembled and how they are dispersed,
tracing out key steps in the cloud-to-star life cycle \cite{2015MNRAS.454..238W,2016AA...587A..76V,2017MNRAS.472.4797S}.
This opens up the possibility to answer 
long-standing questions in galaxy and molecular cloud 
formation that, until Hyperion, have largely been the purview of numerical experiments:
How does
gas in galaxies flow from diffuse-atomic to dense-molecular gas? 
What sets the
size and distribution of molecular clouds in galaxies? 
Is the dominant process for molecular cloud formation
 from large-scale disk gravitational
instability or via phase transitions driven by small scale local compression and/or via cloud collisions into spiral arms?
How does this cycle relate to 
larger scale galactic and intergalactic environments?

Furthermore, while the formation of \hh represents the first
step to star formation, it is unclear how the two processes are
linked. Do molecular clouds evolve in distinct phases, first accumulating
molecules, then forming stars, and
then dissociating because of stellar feedback,
such that the lifetime of an individual \hh molecule
is comparable to that of a MC? Or do molecular clouds
spend most of their 20–30 Myr lives in a steady
state, with \hh formation balancing mass loss by
star formation and \hh destruction with individual \hh molecules
living much less time than molecular clouds (e.g. undergo many cycles of formation and dissociation during the cloud life)? Or does
MC formation and destruction involve mechanical
agglomeration and dispersal of molecular
mass, but with little or no chemical/phase change, so
that individual \hh molecules live much longer
than molecular clouds? 

By providing a \textit{direct measurement of \hh formation
and destruction rates}, Hyperion will answer
these critical questions for the first time.


\subsubsection{Hyperion's Measurements of \hh Dissociation and Formation}

\subsubsection{\hh Dissociation with Hyperion} \label{sec:diss}
We can use measured \hh line emission to estimate the \hh dissociation rate.
The idea is that the total surface brightness summed over the fluorescent line emission is proportional to the \hh dissociation rate, integrated along the LOS.
This is because the \hh lines are excited by FUV LW radiation, which is also responsible for \hh destruction \cite{1989Sternberg}. \hh dissociation measured by Hyperion will occur on the surface of the molecular cloud, where UV photons are easily able to penetrate. While Hyperion cannot probe deeper within the cloud, our primary science is concerned with the destruction of \hh at the cloud boundary. We take care to estimate how much dust attenuation is occurring even at this surface layer.

The complication is that the observed brightness of the emitted lines is reduced compared to the real true excitation rate as the lines suffer from dust absorption as they propagate from the cloud interior to the observer.
 Through measurements of dust extinction, we can estimate the effective dust attenuation and thus how much of the cloud volume do the H$_2$ line observations probe.
 Typically, $\tau \approx 10^{-21} N$ cm$^2$.
Thus, for cloud columns $N < 10^{21}$ cm$^{-2}$ or lower, $\tau < 1$, dust absorption is not significant, i.e., this is the optically thin limit.
For regions in the cloud where $N \gg 10^{21}$ cm$^{-2}$, dust absorption is significant and then the H$_2$ lines do not probe the H$_2$ dissociation along the entire LOS and thus underestimate 
the true integrated dissociation rate.

\subsubsection{\hh Formation with Hyperion}\label{sec:formation}
Measuring \hh formation directly with the \hh florescent spectrum requires a detailed understanding of how the OPR depends on the temperature and ionization state of the region in question.
The imprint of \hh
formation in the cloud is
traced by the OPR, which can be used to measure how far out of equilibrium the \hh is at present.  

 The OPR method is effective in measuring \hh formation time scales at low observed OPR values, where there is no ambiguity as to the state of the gas. Searches for low OPRs indicative of old molecular gas are best conducted in the weak PDRs near cloud edges where the FUV field is dominated by the background interstellar radiation field, rather than toward bright, local sources of dissociating radiation surrounded by intense PDRs. In such weak PDR regions the OPR we measure should be close to that of the bulk chemical state in the cloud interior, since most of the H$_2$ molecules we observe will be pre-existing ones, rather than ones newly re-formed immediately following dissociation. An added benefit is that, in such regions, dust absorption is a minimal concern, since the volume and column densities are likely to be moderate. Once we have measured the chemical age of the H$_2$, we can deduce the approximate H$_2$ formation rate simply by dividing the H$_2$ mass by the chemical age.
 
As noted above, if we measure high OPR values in our target regions, this could mean that the bulk of the molecular gas is truly young, but it could also mean that we have not successfully probed the cloud interior, and we are instead mostly measuring young, surface H$_2$ that is chemically different from the equilibrium gas deeper in the cloud. To distinguish between these two possibilities, and provide an alternative estimate of the H$_2$ formation rate if our measurements find high OPRs, we will employ another method for deducing the \hh formation rate, based on 21cm observations which trace the HI column density. In a forthcoming series of papers (Bialy et al., in prep; Burkhart et al., in prep.) we present the details of this latter method as well as the use of \hh line emissions to trace the \hh dissociation rate, and test these methods against synthetic observations and compare the observed rates to the global star formation rate in the cloud.

\subsubsection{Connecting \hh formation rates and \hh dissociation rates to star formation rates}
Hyperion's measurements of \hh formation and dissociation rates can be combined with measured cloud star formation rates and efficiencies to determine what, if any, relationship exists between the formation of molecules and the global properties of star formation.

\begin{figure*}
    \centering
  \includegraphics[width=\columnwidth]{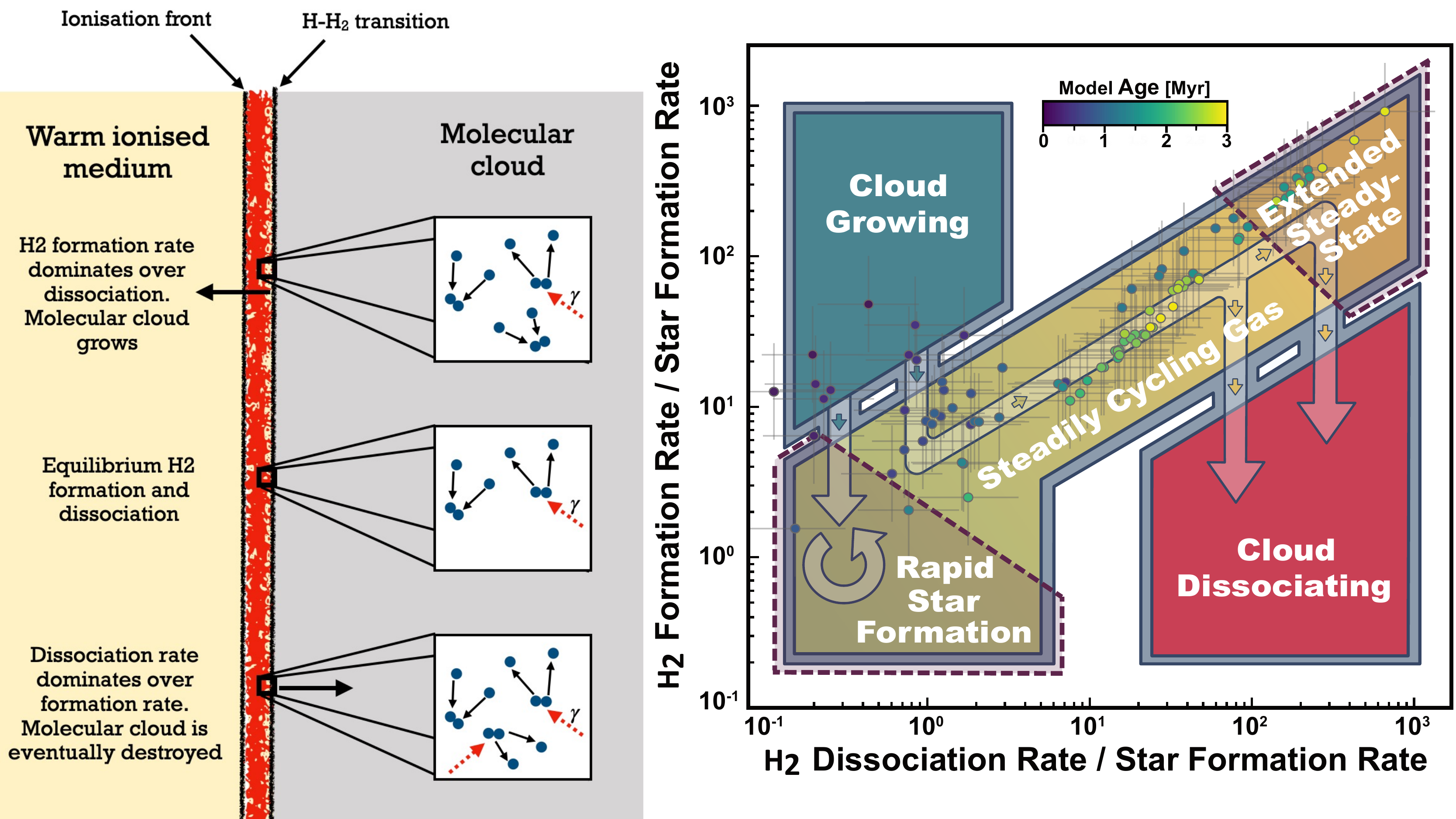}
    \caption{\textbf{Left panel:} Cartoon of the molecular cloud boundary where Hyperion observes. Hyperion is able to determine the rates at which molecular hydrogen molecules are forming and dissociating at the cloud surface, determining if the cloud is growing or shrinking over time. \textbf{Right Panel:} Hyperion compares ratios of \hh formation and dissociation to star formation rate to pinpoint the cloud’s evolutionary state. Four different evolutionary possibilities are shown in different colored area. We show a comparison of simulated points using the Simulating the Life-Cycle of molecular Clouds (SILCC) simulations. Purple and blue points are early evolutionary states (i.e. t $<1.5$ Myrs) while green and yellow points show later evolution but before the cloud is dispersed by supernova $1.5 < t < 3 $ Myrs. Hyperion will empirically sample all evolutionary paths in this plot.}
    \label{fig:star formationr}

\end{figure*}

A cartoon example of the comparison of the cloud star formation to \hh formation and dissociation is shown in Figure  \ref{fig:star formationr}, which shows the ratios of \hh formation to
star formation versus \hh destruction to star formation.  A plot like this is not possible with existing telescopes and is only possible with Hyperion.  Such a diagram demonstrates our ability to disentangle different cloud evolution scenarios discussed above. 

In
a scenario where clouds have distinct phases, they
should be born in the upper left “cloud growing”
region (purple oval), rapidly pass through the 1:1 “steadily cycling
gas” region (yellow oval), and then rapidly move into the
“cloud dissociating” region (pink oval). In the extended
steady-state scenario, where \hh molecule lifetimes
are short, we should see the exact opposite
distribution: most clouds on the 1:1 line, and concentrated
in the upper right “extended steady
state” where \hh formation and destruction rates
greatly exceed star formation rates \cite{2021RNAAS...5..222K,1980ApJ...238..148B}. If MC formation
and dispersal is mainly mechanical and not chemical,
and molecules live much longer than clouds,
molecular clouds should concentrate in the lower left “rapid
star formation” where star formation rates exceed both \hh
formation and destruction. Within Galactic molecular clouds,
different parts of the cloud may appear in different
regions on Figure \ref{fig:star formationr}, although we expect a
high degree of correlation on small scales ($\sim$0.1
pc) \cite{2010ApJ...712L.116P,2015MNRAS.450.4035F}  between formation and destruction
rates due to filamentary structures in molecular clouds.

\subsection{Feedback: Determining how molecular clouds are destroyed}\label{sec:feedback}
Massive stars emit copious amounts of energy into their surroundings in the form of ionizing and dissociating radiation, stellar winds, radiation pressure and eventually supernovae \cite{2014PhR...539...49K, 2015NewAR..68....1D}.  These feedback mechanisms impact the star formation process and the broader interstellar medium in various ways.

Locally, feedback makes star cluster formation a self-regulating process by setting the characteristic mass of the stellar initial mass function \cite{2009MNRAS.392.1363B}. At the scale of the star forming region, feedback acts to disperse the molecular cloud that produces the young stars, limiting further star formation \cite{2015MNRAS.450.1199D, 2015NewAR..68....1D, 2017MNRAS.464.3536R}. However, feedback can potentially trigger new star formation in the same cloud and even neighbouring clouds \cite{2005AA...433..565D, 2006AA...446..171Z, 2007AA...472..835Z, 2008ApJ...688.1142K, 2009AA...497..789U, 2009ApJ...694L..26G, 2011ApJ...736..142B, 2011MNRAS.412.2079M, 2012MNRAS.420..562H, 2012MNRAS.421..408T, 2012AA...546A..33T}. At smaller scales, feedback can heat and disperse circumstellar disks around young stellar objects, potentially impacting the planet formation process (discussed further in Sec \ref{sec:diskdispersal}). 

At much larger scales, feedback impacts the interstellar medium through supernova explosions. Supernovae are capable of injecting energy out to much larger distances which is important for regulating the gas distribution in the interstellar medium and hence properties such as the galactic scale height and the mean star formation rate \cite{2011MNRAS.417.1318D, 2015MNRAS.447.2144D}. Indeed calibrated ``feedback'' is a key sub-grid process for models of the formation and evolution of galaxies, which are some of the worlds biggest astrophysics simulations \cite{2014MNRAS.445..175G, 2014Natur.509..177V, 2017PhRvL.118p1103L}. However as a sub-grid process the nature of the feedback itself is not really understood in these models, with the prescription tuned to give results in agreement with observations. Recent work has shown that the impact of supernova feedback depends on preprocessing work by radiation and winds. A larger volume of the interstellar medium is influenced by supernova feedback if winds and photoionization sculpt low density channels in the star forming cloud before the supernova goes off,  \cite{2013MNRAS.431.1337R, 2020MNRAS.493.4700L}. Therefore, a better understanding of the preprocessing role of feedback, which is one of Hyperion's goals, is crucial to connecting the scale of star-forming regions and that of the wider Galaxy, though recent theoretical efforts such as the SILCC project \cite{2016MNRAS.456.3432G, 2017MNRAS.466.3293P, 2017MNRAS.472.4797S, 2019MNRAS.482.4062H, 2021MNRAS.504.1039R} and other groups \cite{2020MNRAS.495.1672B, 2022MNRAS.510.5592A, 2022MNRAS.509.6155R} are working to address that. 

For some time, photoionization has been expected to be the dominant feedback mechanism prior to supernovae \cite{2014MNRAS.442..694D, 2020MNRAS.492..915G}. However, recent observations in [C II] as part of the SOFIA FEEDBACK survey  have found new evidence for fast flows driven by winds, which they suggest implies a more dominant role of winds \cite{2019Natur.565..618P, 2020AA...639A...2P, 2021SciA....7.9511L}. Whether these fast flows truly represent a widespread wind that is a key driver of the cloud dispersal or they only trace a handful of low density channels through which the wind efficiently escapes remains an alive debate in literature \cite{2020MNRAS.492..915G, 2021MNRAS.501.1352G}. 

Given the importance of feedback discussed above, it is necessary to: \\

\noindent i) Observationally determine the relative impact of different feedback mechanisms in dispersing molecular gas over time. \\

\noindent ii) Observationally determine whether supernovae in a region will be confined and deposit their energy locally, or whether they will be able to inject energy into larger scales. \\

But why has this been difficult to achieve to date, and how will Hyperion succeed? 

Determining the feedback mechanism responsible for dispersal requires an estimate of the energy output from massive stars and the kinematics of the surrounding gas. The issue with the kinematics is that widely used tracers such as CO are downstream of the feedback in the neutral/outer PDR part of the flow and tracers in the H\,\textsc{ii} region generally don't probe the rate of expansion. This is why the recent [C II] SOFIA observations in the FEEDBACK survey have been so impactful and surprising. They have enabled us to probe the kinematics of components of the feedback driven material that we previously had no access to. What we ideally need is a tracer at the driving interface, in the vicinity of the ionization front and H-H$_2$ transition, to properly trace the expansion kinematics. UV fluorescence emanating close to the H-H$_2$ transition offers this probe. Due to the high optical depth of the ISM to the UV, Hyperion will primarily trace receding accelerated surfaces through the large holes in molecular gas associated with HII regions. Hyperion will use this to measure the expansion of feedback driven surfaces in H\textsc{ii} regions and determine whether they are consistent with thermal expansion, winds or supernova driving based on the velocity of the expansion for size of the H\textsc{ii} region. Due to the high optical depth of the ISM to the UV, Hyperion will primarily trace receding accelerated surfaces through the large holes in molecular gas associated with HII regions

\begin{figure*}
    \centering
    \includegraphics[width=\columnwidth]{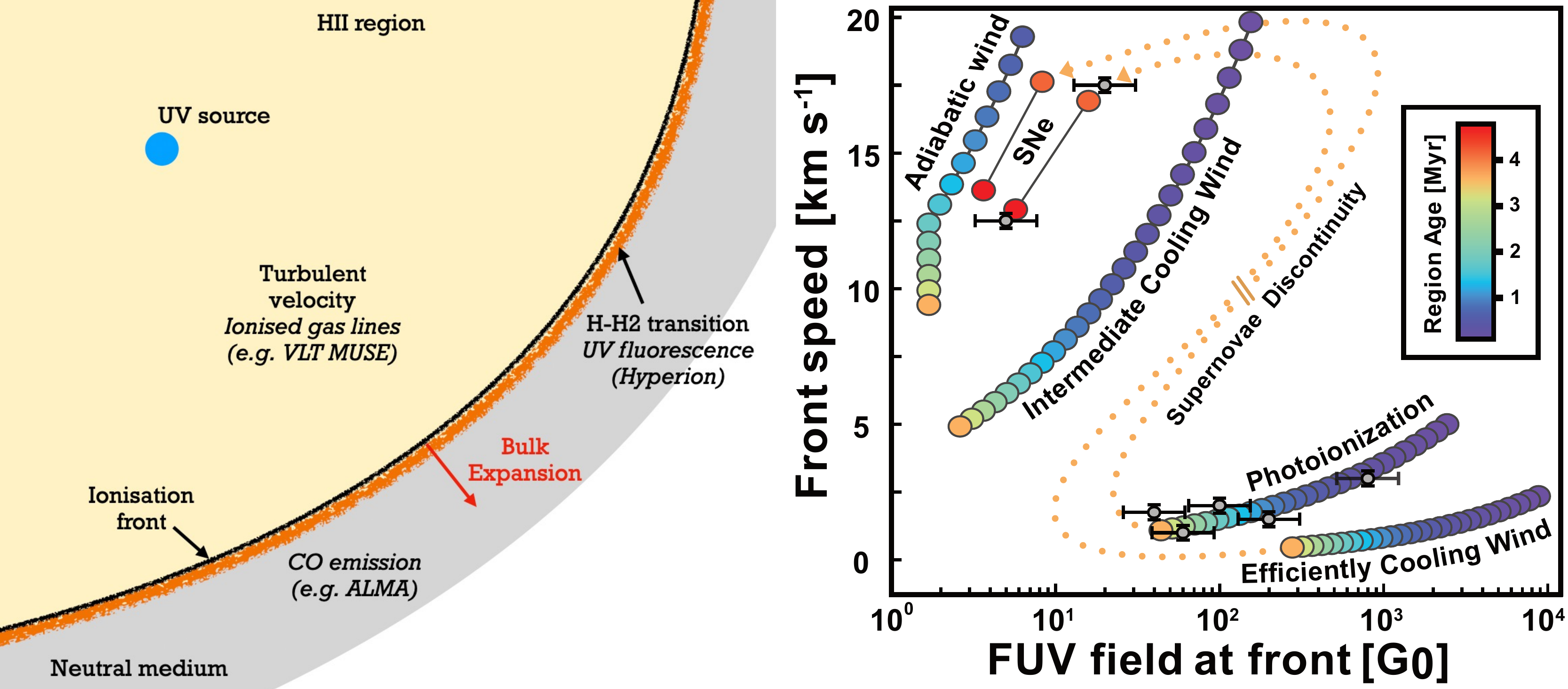}
    \caption{\textbf{Left Panel:} Cartoon of a bright UV source blowing a bubble in a molecular cloud. Hyperion observes the edges of the bubble, at the HI-\hh transition, to determine the expansion speed of the front and the driving mechanism of the bubble. \textbf{Right Panel:} Hyperion distinguishes
mechanisms behind feedback-driven bubbles by measuring their expansion speeds and FUV radiation fields. Colored dots are semi-analytic models. Points with 1-$\sigma$ errors are synthetic Hyperion observations of models.}
    \label{fig:hyperionSO2cartoon}
\end{figure*}

A valuable characteristic of the H$_2$ fluorescent spectrum is that the intensity of certain lines is a function of the exciting FUV radiation field strength \cite{1989Sternberg, 1997ApJ...475..835W}. The combination of the FUV field at the expanding front and the expansion velocity can be used to distinguish the driving feedback mechanisms, as illustrated in Figure \ref{fig:hyperionSO2cartoon}. This figure was constructed using semi-analytic models of expanding H\,\textsc{ii} regions \cite{2015MNRAS.453.1324B, 2018MNRAS.479.2016W} and and wind blown bubbles\cite{1977ApJ...218..377W}.

In addition, the long slit of Hyperion can be used to map the radial profile of the FUV radiation field by measuring line intensities as a function of the distance to the FUV source near the bubble center. A line of sight through which supernova energy can escape to larger distances will also be associated with a radial FUV profile without an abrupt decrease associated with a dense molecular wall. By sampling the angular profile of the FUV radiation field we can determine the fraction of angles through which radiation is escaping to larger distances and hence determine if the region is predominantly radiation bounded (in which case the supernova energy deposition will be local) or otherwise. Achieving this requires being able to map the radial profile of the exciting UV out to sufficiently large distances from the UV source. Given that the H$_2$ UV fluorescent emission intensity is an increasing function of the strength of the exciting UV radiation, the high spectral resolution high sensitivity design of Hyperion is necessary to achieve this.

Hyperion maps the projected escape of FUV radiation, but given that the holes in H\textsc{ii} regions are typically large (rather than a large number of small holes like a sieve) the projected fraction of angles on the sky (i.e. fraction of $2\pi$ in projection) associated with low density channels correlates with the fraction of $4\,\pi$ steradians through which supernova energy can efficiently escape to larger distances. Hyperion will map the FUV profiles of star forming regions, which following the above allows us to quantitatively constrain the dense cloud coverage and consequences for supernova feedback. 

The above applied to multiple massive star forming regions will combine to tell us what is driving the pre-supernova feedback responsible for displacing the bulk of the molecular gas, and the impact this has in enabling supernova feedback to escape into the wider ISM. This will provide a vital link between that of the star forming region and galaxy evolution.

\subsection{Disks Dispersal: Evaluating the mechanisms for gas disk dispersal}\label{sec:disks}
\label{sec:diskdispersal}
It is now observationally well established that exoplanets are remarkably diverse in their properties \cite{2015ARAA..53..409W}. Determining the origin of this requires understanding the formation and evolution of planetary systems, which in turn requires understanding the evolution and dispersal of planet-forming disks. 

Disk dispersal may proceed in a number of different ways: \\

\noindent i) The transport of material through the disk, resulting in accretion onto the star \cite{2016AA...591L...3M, 2020AA...639A..58M, 2021AA...650A.196M}.  Until relatively recently, this transport was thought to be facilitated by a turbulent viscosity \cite{Shak}, though recent observations do not provide firm evidence for the required degree of turbulence. There is growing evidence for the required transport of angular momentum being facilitated by magnetohydrodynamically driven winds (MHD winds; see point iii below).  \\

\noindent ii) The growth and radial drift of dust grains through the disk and onto the host star \cite{2012AA...539A.148B}. Without pressure bumps in the disk to trap dust \cite{2012AA...538A.114P, 2018ApJ...869L..41A}, this can lead to the very rapid depletion of the dust budget on short timescales.    \\

\noindent iii) The extraction of material through MHD driven winds. MHD winds can be launched as a result of the magneto-rotational instability \cite{1991ApJ...376..214B, 2009ApJ...691L..49S, 2016AA...596A..74S} (MRI) or magnetocentrifugally \cite{2007prpl.conf..277P, 2014prpl.conf..411T, 2016ApJ...818..152B, 2020ApJ...896..126G}. The aforementioned work showed that, in addition to extracting material, MHD driven winds can also efficiently redistribute or remove angular momentum, facilitating accretion.    \\

\noindent iv) The extraction of material in internal photoevaporative winds. X-ray, EUV and FUV radiation from the host star can heat material in the inner disk above the escape velocity, driving a photoevaporative wind \cite{2009ApJ...690.1539G, 2010MNRAS.401.1415O, 2011MNRAS.412...13O, 2012MNRAS.422.1880O, 2015ApJ...804...29G, 2017RSOS....470114E, 2019MNRAS.490.5596W, 2019MNRAS.487..691P, 2021MNRAS.508.3611P}.  

\noindent v) The extraction of material in external photoevaporative winds. Material in the outer disk is less strongly bound and so can be driven in a wind by more modest heating than the inner disk. The FUV radiation in a wide range of star forming environments can be sufficient to drive material from the outer disk \cite{1993ApJ...410..696O, 1994ApJ...436..194O, 2000ApJ...539..258R, 2004ApJ...611..360A, 2016ApJ...826L..15K, 2017AJ....153..240A, 2018ApJ...860...77E, 2018MNRAS.481..452H, 2019MNRAS.485.3895H, 2021MNRAS.501.3502H}. This is expected to reduce the disk mass, radius and lifetime \cite{2018MNRAS.478.2700W, 2019MNRAS.490.5678C, 2020MNRAS.492.1279S, 2022MNRAS.512.3788Q} and could also influence the migration and mass of massive planets in the outer disk (Winter et al. in preparation). It is currently thought that the majority of young stars form in an environment that would lead to significant external photoevaporation \cite{2008ApJ...675.1361F, 2020MNRAS.491..903W}, however the nearest star forming regions and hence sites of our best studied discs are low mass such as Taurus/Lupus with an absence of strong UV sources \cite{2018ApJ...869L..41A, 2021ApJS..257....1O}.   \\

\begin{figure*}
    \centering
    \includegraphics[width=0.5\columnwidth]{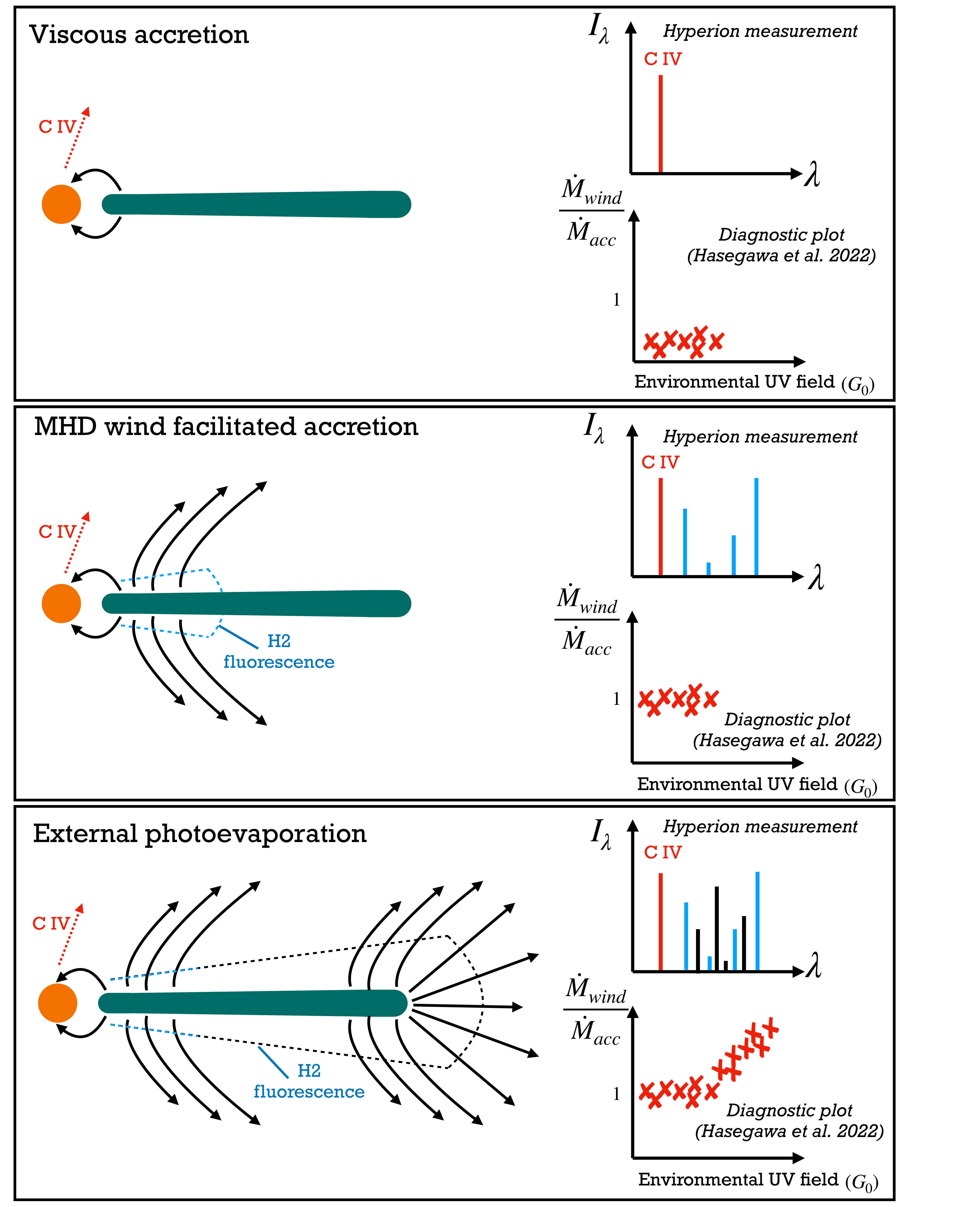} 
    \caption{A cartoon illustrating the Hyperion measurements of protoplanetary disks dispersing by different pathways and the diagnostics that will be used to identify each regime. }
    \label{fig:hyperionSO3cartoon}
\end{figure*}

\begin{figure*}
    \centering
     \includegraphics[width=0.5\columnwidth]{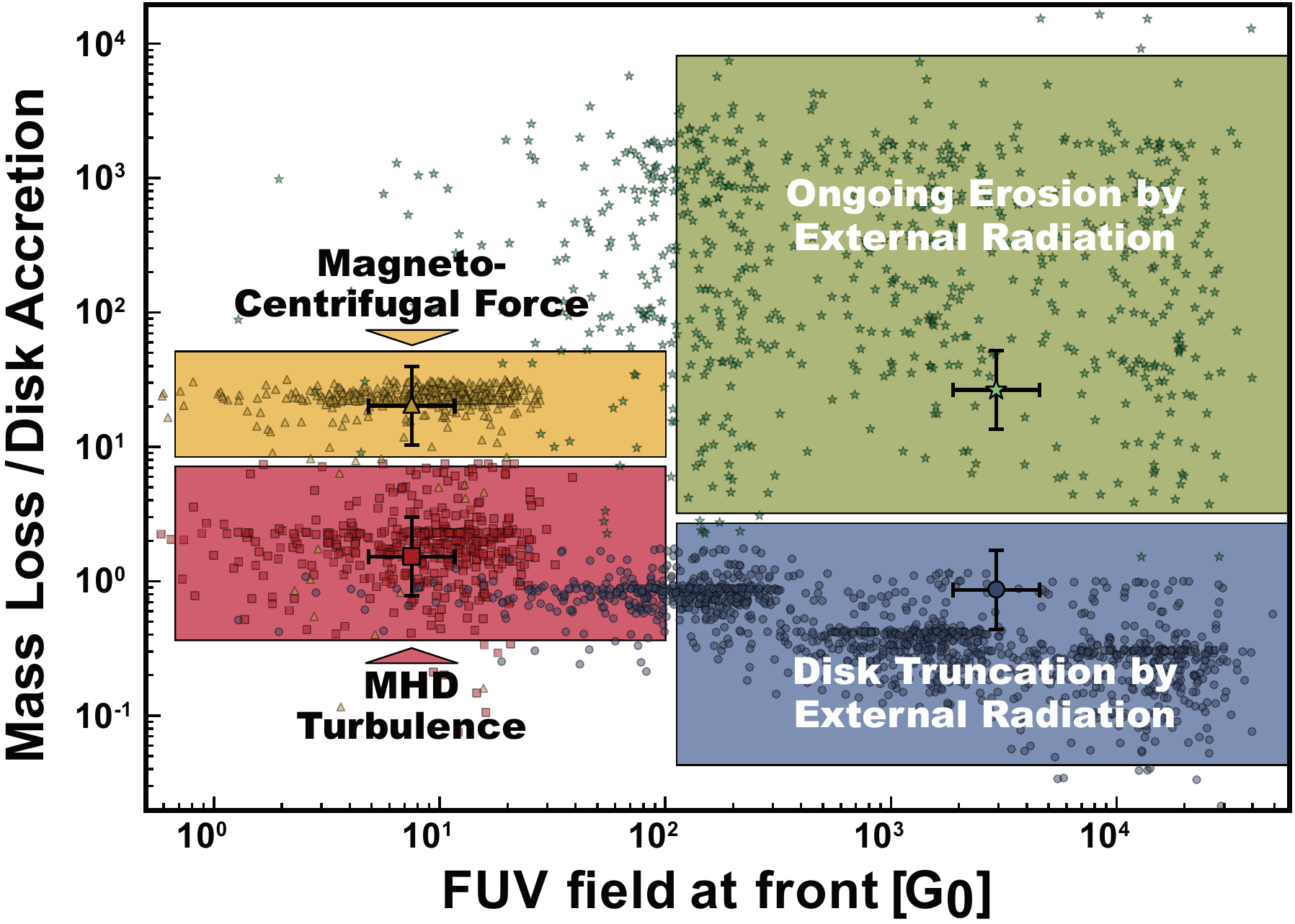}
    \caption{Hyperion will determine under what conditions planet-forming disks undergo mass accretion and dispersal through winds driven by several mechanisms (colored regions and points). Points indicate simulated observations by Hyperion.}
    \label{fig:hyperionSO3keystone}
\end{figure*}

Each of the above mechanisms have been and continue to be studied in detail. However, to date, their full interplay is too complicated to study theoretically, while observations can only probe a very limited subset of processes at any given time. For example, there are separate tracers of the gas and dust components of disks \cite{2018ApJ...869L..41A, 2021ApJS..257....1O}, accretion \cite{2016AA...591L...3M, 2020AA...639A..58M, 2021AA...650A.196M}, inner winds \cite{2008MNRAS.391L..64A, 2011ApJ...736...13P, 2014AA...569A...5N, 2016ApJ...831..169S, 2019ApJ...870...76B, 2020MNRAS.496.2932B} and outer winds \cite{1993ApJ...410..696O,1998ApJ...502L..71S, Henney99, 2000ApJ...539..258R,  2020MNRAS.492.5030H}. Semi-analytic theoretical models suggest that the interplay between accretion and internal and external winds provides the key controlling mechanism for disk evolution \cite{Hasegawa2022, 2022arXiv220402303C}; however, no single instrument can probe all of these. Hyperion offers the ability to do so for the first time. 

Hyperion is capable of simultaneously measuring both internal and external winds from disks through UV-H$_2$ fluorescence and the relative level of accretion using the C\,IV 1549\AA\ emission doublet, as shown in Figure \ref{fig:hyperionSO3cartoon}. The internal and external winds are not spatially resolved by Hyperion. Rather, it distinguishes them using the fact that specific emission lines arise in the fluorescent spectrum, depending on the source of H$_2$ excitation. In addition, the intensity of H$_2$ emission is a function of the incident FUV field strength, and the outer wind is generally driven by a weaker FUV radiation field than the inner component, leading to fainter emission lines in the outer wind. Furthermore, the flow speed is estimated by velocity centroiding the H$_2$ lines, which provides further distinction between the slow ($\sim$4 km\,s$^{-1}$) outer winds and faster (tens of km\,s$^{-1}$) inner winds. 

Hyperion's measurements will enable it to compare the rate of mass loss through winds with the accretion rate, which will allow it to determine whether the internal winds are sufficient to drive accretion and discover whether additional mass loss due through external photoevaporation plays a significant role in protoplanetary disk dispersion, and if so, in what kinds of external UV radiation environments \cite{Hasegawa2022}. A cartoon of how to distinguish between these is shown in Figure \ref{fig:hyperionSO3keystone}.

\section{The Hyperion instrument design}\label{sec:instrument}

\begin{figure}
  \includegraphics[width=0.5\columnwidth]{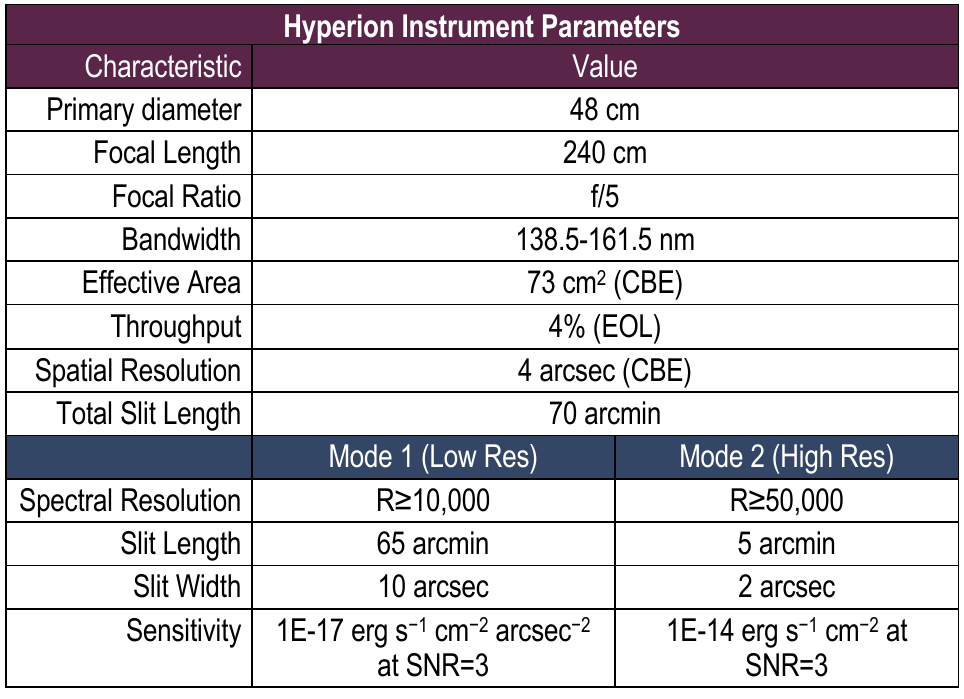}
    \caption{Overview of Hyperion Instrument Parameters. EOL=End of Life, CBE= Current Best Estimate.}
    \label{Tab:Instrument}
\end{figure}

To achieve the science objectives outlined in this paper, we have designed a single instrument that provides a balance between spectral and spatial resolution, design complexity, and cost. Briefly, Hyperion consists of a 48 cm Cassegrain telescope, focusing light onto a 70 arcmin long slit, which feeds an Offner style spectrometer. The spectra are captured by a linear mosaic of 4 high efficiency delta-doped CCDs, optimized to the Hyperion bandpass. All optical surfaces are reflective, with a high heritage aluminum coating protected with MgF$_2$. At the Hyperion bandpass, this provides reflectance above 90\%, even assuming typical contamination at end of life of 3\% (based on HST performance over time, \cite{2010Quijada}. The overall Hyperion instrument design parameters are listed in Figure \ref{Tab:Instrument}.

\begin{figure}
\centering
  \includegraphics[width=0.5\columnwidth]{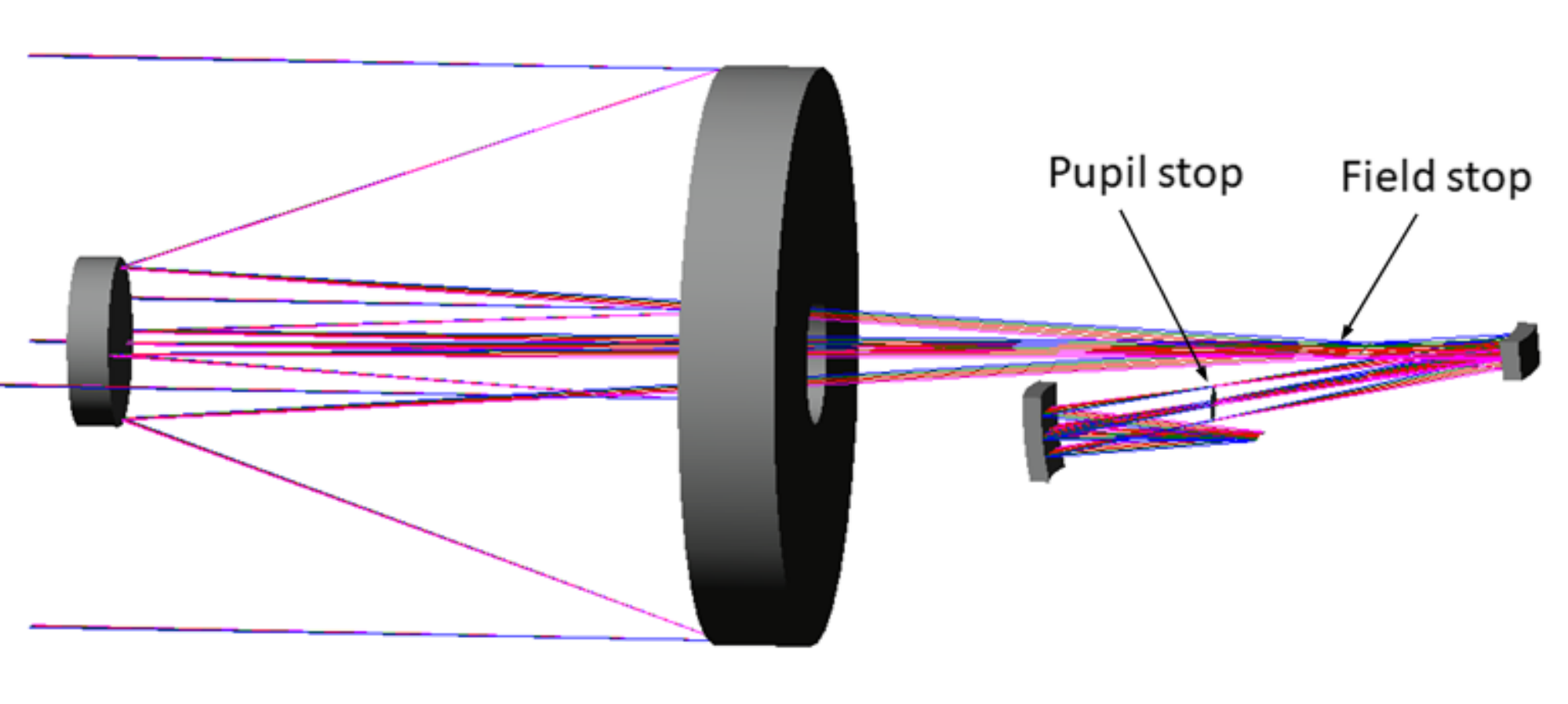}
    \caption{Zemax raytrace of the Hyperion telescope optics, including locations of the pupil stop and field stop.}
    \label{fig:layout}
\end{figure}

The Hyperion telescope optical layout is shown in Figure \ref{fig:layout}. The telescope design is an on-axis 48-cm Cassegrain with a two-element corrector with tilted and decentered free form surfaces. Performance readily supports Hyperion spatial resolution, which is SNR driven (it does not need to be diffraction limited).

The long length of the slit (70 arcmin total length) is driven by needing to cover as large an angular area at moderate resolution as possible in order to map molecular hydrogen fluorescence over the large fields of view occupied by nearby Galactic molecular clouds. Achieving good performance over this length requires the addition of a 2-mirror relay (with above 90\% reflectance) between the Cassegrain telescope and the slit to keep the FWHM of the PSF to between 4\arcsec-7\arcsec\ over the 70\arcmin\ field. The scientific observation throughput gain from the wide field of view surpasses the moderate loss from the two extra reflections. The relay also provides an intermediate image plane and Lyot stop (pupil plane) to suppress straylight. 

Hyperion's stepped slit design provides two different spectral resolutions, enabling two observing modes without any moving parts. The majority of the slit length is 10 arcsec wide, providing a resolution of R$\sim$10,000 over 65 arcmin. The central 5' portion of the slit is 2 arcsec wide, giving a high spectral resolution of 50,000 or greater. This geometrically shaped slit design reduces complexity (no moving parts) while still providing flexibility to achieve all of Hyperion's science objectives.

The Hyperion spectrometer consists of a collimating mirror, a grating, a camera mirror, and a fold mirror before the focal plane. We choose to use the grating in the third order to simplify fabrication. This increases the period to a manufacturable 330 nm. The efficiency fall-off with wavelength is manageable because of the narrow Hyperion band pass. Zeiss has made holographic gratings of comparable sizes and periods and also has experience with convex substrates. The efficiency in the system model were based on profiles achieved by Zeiss on previous gratings.

The Hyperion focal plane consists of 4$\times$1 mosaic of delta-doped UV optimized CCDs, which are passively cooled by a radiator to $<$ 168K, and is then actively controlled with a closed-loop heater circuit at 168 K. Each detector is a Teledyne e2v quad readout CCD272-84, which is derived from the CCD273 used on the European Space Agency's (ESA) Euclid mission. The CCD has a 4k $\times$ 4k format with 12 $\mu$m pixels. The detectors are UV optimized via delta-doping \cite{2017Nikzad,2016Hamden,2012Nikzad,2014Hoenk} and a detector-integrated, directly-deposited metal-dielectric filter (MDF) \cite{Hennessy2015,2017Hennessy} which provides out of band rejection. The UV camera system meets all Hyperion performance and environmental requirements, and has been demonstrated to be TRL 6 by means of a combination of analysis and measurements on high fidelity prototypes. Historically, UV missions with microchannel plate detectors (MCPs) have not been able to observe in the UV bright regions that Hyperion investigates because of their count rate limits. As a brief example, Hyperion's observation plan of the Orion region includes the bright star iota Orionis. This star has a measured flux at 1565 of 8.483 $\times$ 10$^9$ erg cm$^{-2}$ s$^{-1}$ angstrom$^{-1}$. Using Hyperion's bandpass and effective area, we find an expected count rate from this star alone to be around 6M counts/s. In contrast, the Europa-UVS mission uses an MCP with a global count rate limit of 1.2 M counts/s \cite{2022Davis}. Hyperion's targets exceed these limits.

Hyperion's ability to detect faint sources requires long exposures to meet the sensitivity requirements at suitable SNR. The Hyperion instrument is read noise dominated preventing the use of arbitrarily many shorter exposures. Long exposures bring challenges with pixels being knocked out by cosmic rays. For each objective, there is an optimization of exposure number, duration, and on-chip binning to achieve the required SNR and ensure only 0.3-1.4\% of bins are lost to cosmic ray hits. Exposure durations range between 930 and 5000 s (16 to 66 mins) depending on target brightness and observing mode.

\section{Hyperion spacecraft and payload overview}\label{sec:spacecraft}

The Hyperion spacecraft is based on the BCPSmall product line from Ball Aerospace. Hyperion benefits from the successful on-orbit NASA GPIM and WISE along with USAF STPSat-2 and STPSat-3 missions. The IXPE, SPHEREx, and
SWFO-L1 missions are currently under development for NASA using the BCP-Small S/C, with SPHEREx providing the most similar mechanical configuration to Hyperion.

The passively-cooled payload is protected from the sun with a Kepler-style wrap-around sunshield enabling the science-required $\pm30^{\circ}$ boresight roll (to ensure any position angle on a target over the course of a year). Fixed solar arrays are mounted to the sunshield. The average science-mode sun incidence is a $9^{\circ}$ roll and $36^{\circ}$ tilt resulting in a 31\% EOL power margin. The worst-case sun angle is $80^{\circ}$ tilt with no roll leading to use of the batteries. Depth of discharge never exceeds 40\% during observing (batteries are sized for infrequent but long duration  eclipses: ~5 hours 3 times a year).

Achieving a lunar-resonant orbit requires a lunar flyby, mission orbit insertion burn and a number of deterministic and statistical trajectory correction maneuvers. These vary in number and duration depending on the day within the 21-day launch period. The maximum required $\Delta$V is 193 m/s including maneuver contingency, operational desaturation, and uncertainties and residuals. No stationkeeping is required and the Hyperion science orbit exhibits excellent long-term stability, maintaining a perigee altitude
well above the geosynchronous regime for at least 100 years. At end-of-mission, the S/C will be passivated but no maneuvering will be required to ensure planetary protection compliance. The tank will be loaded with 113kg of propellant (340 m/s total $\Delta$V for a 554 kg observatory dry mass), a 60\% propellant margin which can support either maneuver uncertainties or an extended mission.

The observatory has comfortable margins in launch mass, power, link budget, data storage, slew speed and pointing accuracy (requirement: 95\% probability that a point source will be found $\pm$5\arcsec\ perpendicular to the slit). The principal challenge for this mission is pointing stability sufficient to ensure a 95\% probability that a point source remains within the 2\arcsec\  slit throughout an exposure of up to 5000 s. This can be achieved (with 30\% contingency and 25\% margin) with the use of three optical head star trackers mounted on the anti-sun side of the spacecraft, and benefiting from the stable thermal environment of an orbit so far from Earth.

\section{Hyperion's orbit, observing plan, and survey strategy}\label{sec:survey}

Hyperion has a single instrument with a single observing mode -- point-and-stare, long-slit spectroscopy. Each science objective has its own strategy for utilizing the instrument via optimal slit placement. To efficiently conduct long exposure observations on faint sources, and to reduce the impact of radiation from the Van Allen belts, Hyperion resides in a TESS-like Lunar-resonant orbit. Low Earth orbit (LEO), in contrast, can accommodate exposures of only up to $\sim$30 minutes in most directions. In a TESS-like orbit, objects are continuously visible for nearly two weeks and the maximum exposure time is only limited by the cosmic ray rates. In addition, most LEO's have regular periods of eclipse by the Earth's shadow, yielding a challenging thermal environment. Other orbits that were studied are an intermediate, highly-elliptical orbit, which would result in passes through the inner Van Allen belt twice per period, and an L2 orbit, which offers scant additional benefits for observing, but has higher hardware costs. The TESS-like two-week 2:1 Lunar resonant orbit provides a stable orbit with apogee of 400,000 km, no station-keeping, high-thermal stability, and an L2-like radiation environment. Orbit right ascension of the ascending node and inclination are set by the constraint to ensure Orion is visible for at least 5 months of the year.

\begin{figure}
\centering
  \includegraphics[width=\columnwidth]{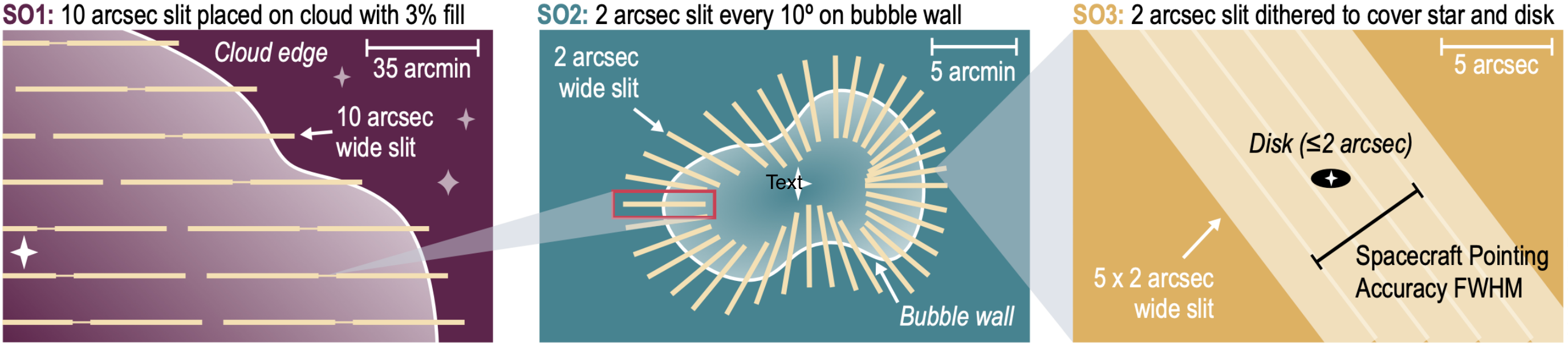}
    \caption{Figure showing the observing scenario for each science objective.}
    \label{fig:survey}
\end{figure}

\begin{figure}
  \includegraphics[width=0.5\columnwidth]{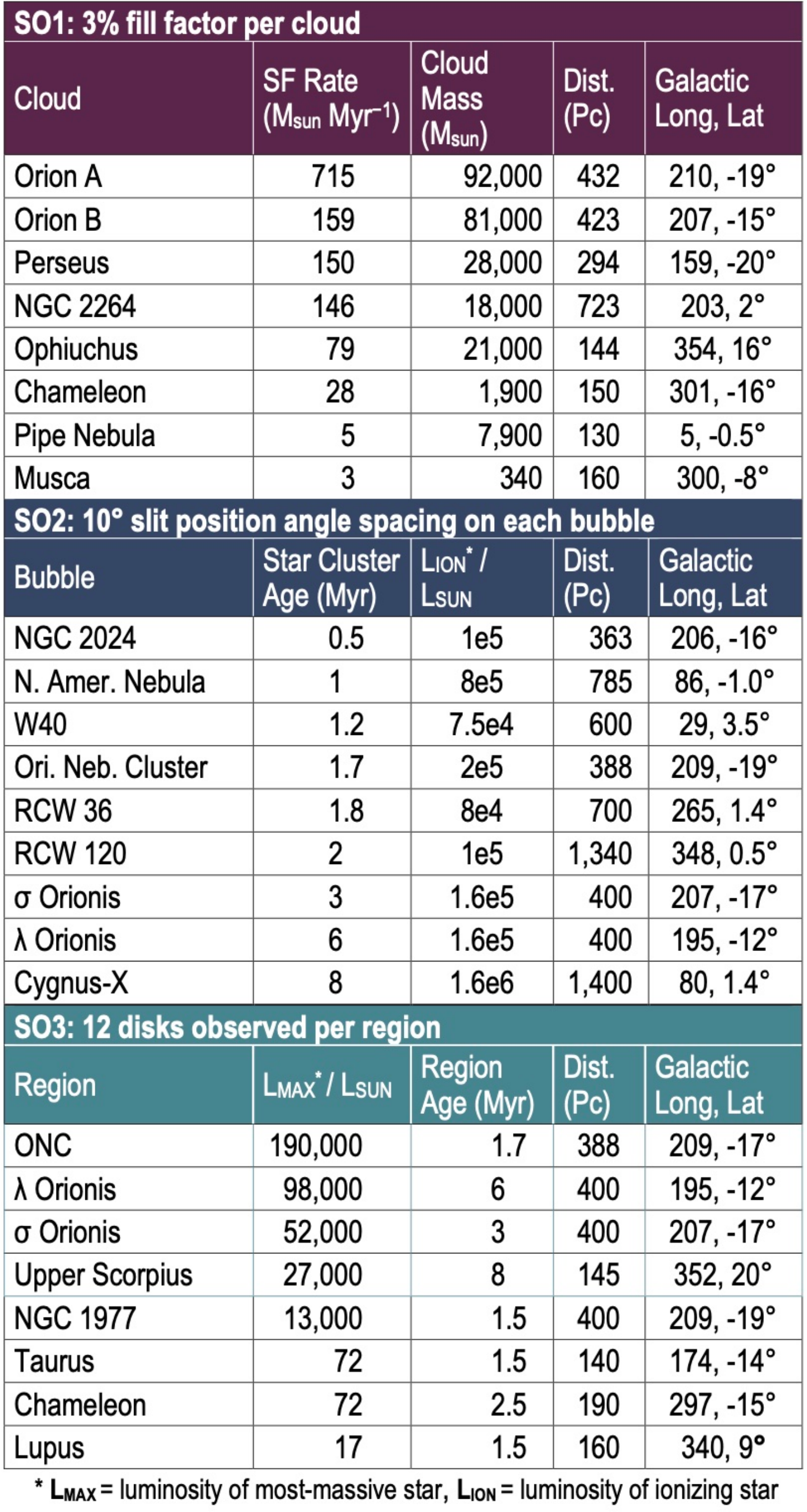}
    \caption{Target list by science objective}
    \label{Tab:Targets}
\end{figure}

The overall observing program consists of an 18 month science phase, split between the three science objectives. All surveys for each objective are completed within the first year, while the final 6 months provide 50\% observing time margin. The observing schedule is constrained mainly by the solar exclusion angle (95$^{\circ}$), with additional requirements set by the Moon and Earth exclusion angles. The strongest science constraint on the observing plan is coverage of the Orion complex, the nearest high-mass-star-forming cloud, whose intense UV environments provide valuable targets for all three of the mission's science objectives. The Orion complex is visible over 5 months of the year from our planned orbit. To be robust to launch dates and adjustments in observing requirements, the Hyperion survey plan has 113\% science margin and completes the baseline mission in a single year, while allocating the remaining 6 months for revisits or follow-up studies. The observing schedule for the targets in Figure \ref{Tab:Targets} is constructed accounting for the Sun, Earth, and Moon pointing constraints and using exposure durations that reach the required SNRs for the expected source brightnesses with the instrument operating at its expected performance.  One-tenth of the time on orbit is reserved for engineering activities including telecoms, slewing, calibration, and decontamination.  On this schedule, the baseline observations are completed within 12 months. Figure \ref{fig:survey}, showing how targets are covered for each science objective. For science objective 1, slits are placed across the targeted molecular clouds to provide uniform $\ge$ 3\% area coverage. This coverage yields a 5-10\% error on the true underlying dissociation rates, based on analysis of the simulations. For science objective 2, the 2\arcsec-wide portion of the slits is placed at spoke angles $<$10$^{\circ}$ apart to cover 360$^{\circ}$ of the bubble wall. For science objective 3, each compact disk is caught by five dithered pointings, offset by 2\arcsec\ to cover the point spread function (PSF). 

During the science mission, Hyperion only requires propellant for momentum dumping (station-keeping and disposal maneuvers are unnecessary). If orbit entry is achieved within the margined propellant budget, Hyperion carries consumables for more than 30 years on orbit. In addition, the expected hardware lifetime is longer than the 18 month science mission. Possible science programs in an extended mission phase are described below.

\section{Data Sufficiency}

Hyperion collects spectra over wavelengths 138.5–161.5~nm to capture enough of the H$_2$ fluorescent lines to reach all three SOs.  The spectra from the 10$^{\prime\prime}$-wide portion of the slit have R=10,000 to separate the stronger lines, including those of the molecule's ortho and para spin states, while the spectra from the slit's narrower, 2$^{\prime\prime}$-wide central portion have R=50,000 to measure the lines' Doppler shifts and widths in bubbles and disks. Each molecular cloud target covers many degrees in area, but we are also interested in the expected small scale features within each giant cloud. The angular resolution along the slit of 7$^{\prime\prime}$ is sufficient to resolve features in bubbles and clouds within a few hundred parsecs and to distinguish target stars from their neighbors.
 
The 10$^{\prime\prime}$-wide portion of the slit is 65$^\prime$ long, enough to survey 3\% of the target clouds' area in the mission lifetime, while the 2$^{\prime\prime}$-wide portion is 5$^\prime$ long, enough to survey the bubbles at the higher spectral resolution.
 
The spectra exceed SNR=3 at the threshold surface brightness of 1E-17 erg s$^{-1}$ cm$^{-2}$ arcsec$^{-2}$ for both the lower and higher spectral resolutions, so the fluorescence is measured down to the level excited by the mean interstellar UV radiation field.  The corresponding sensitivity threshold for the compact sources of SO3 is 1E-14 erg s$^{-1}$ cm$^{-2}$.  Reaching these sensitivity thresholds requires exposures longer than feasible in low Earth orbit.  These are enabled by Hyperion's lunar-resonant orbit.
 
Hyperion surveys the fluorescence in eight clouds forming stars at rates 3 to 715 M$_\odot$ Myr$^{-1}$ (Figure \ref{Tab:Targets}), spanning the range found in our Galactic neighborhood.  The baseline mission maps more than twice the minimum fraction of the eight target clouds' area to determine their main evolutionary pathway on Figure \ref{fig:star formationr}.
 
Hyperion also surveys nine bubbles with ages 0.5 to 8~Myr (Figure \ref{Tab:Targets}), enabling testing the full spectrum of potential dispersal mechanisms (Figure \ref{fig:hyperionSO2cartoon}).  The baseline mission returns exposures at slit angles spaced around the bubbles twice as finely as the 10$^\circ$ spacing below which the leakiness can be gauged.  The spacecraft is capable of placing the slit at any orientation on a given target over a 12-month period, so that all bubble features can be sampled.
 
Finally, Hyperion determines wind mass-loss rates from disks in each of eight star-forming regions lit by the full local range of UV environments, from the average interstellar field to the intensity of the Orion Nebula Cluster (Figure \ref{Tab:Targets}).  Fluxes are measured to a precision of 10\%, so the disk winds' masses and flow rates can be determined as SO3 requires.  The baseline mission measures 24 disks in each star-forming region, twice the minimum to distinguish driving mechanisms (Figure \ref{fig:hyperionSO3keystone}).  Hyperion completes its baseline mission in 18~months.

\section{Other critical science Hyperion can conduct}\label{sec:other}

Hyperion’s observational capability can fuel an active guest investigator (GI) Program after the primary mission ends. This includes observing interstellar gas and star forming regions in low metallicity targets such as the SMC and I Zw 18. Local metal-poor galaxies are ideal analogues of primordial galaxies with a barely enriched ISM. Searches for molecular gas in metal-poor galaxies \cite{2017Oey,2021Zhou,2017Schruba,2015Rubio} are actively ongoing, as the impact of low metal abundance on star formation and \hh formation remains unknown and of great theoretical interest \cite{2013Krumholz}. 

Additional Galactic science includes studying emission lines from shocked and photoionized warm-hot gas in supernova remnants, planetary nebulae, and the ISM. The brightest rest-frame FUV nebular diagnostic lines in the Hyperion window are: CIV 1447, 1551, HeII 1640, NIV] 1483, 1486, NeIV 1600. Possible Galactic investigations include the study of emission lines from shocked and photoionized warm-hot gas in supernova remnants, planetary nebulae, star forming regions, and the ISM. In particular, Hyperion's feedback objective simultaneously provide exquisite data on the exciting FUV field strength, attenuation and $H_2$ distribution of PDRs. Beyond the Milky Way, Hyperion can study stellar populations and emission line regions in nearby galaxies near sites of intense star formation or Active Galactic Nucleus (AGN). 

Beyond the Milky Way, Hyperion can study stellar populations and emission line regions in nearby galaxies near sites of intense star formation or AGN. Hyperion can also probe the circumgalactic medium (CGM) of distant galaxies, observing redshifted SiIV1400, OIV]1404 (z$>$0.0035), CII1335 (z$>$0.05) and H-Ly$\alpha$1216 (z$<$0.15), all important FUV diagnostic signatures. Hyperion can study the impact of AGN and supernova feedback in the z$\sim$0.1 Ly$\alpha$ forest \cite{2017Gurvich,2017Viel} and will be the only observatory in the near-term future after HST COS that can study the properties of the z$\sim$0.1 Ly$\alpha$ forest. Hyperion fills an important gap in UV spectroscopic coverage.

\section{Conclusion}

Hyperion uses the UV fluorescence of molecular hydrogen to trace key phases in the formations and lifetimes of molecular clouds, stars, and planets. Hyperion speaks to key Astro2020 Decadal Survey \cite{2021Decadal} science priorities: \textit{Understand how galaxies evolve through gas condensing to form stars.}; and \textit{Trace stars and planets from the disks out of which they form to the vast array of extra- solar planetary systems.} Hyperion does this by achieving 3 science objectives which explore how MCs and disks lose or gain mass as a function of their surroundings and internal actions. 

Hyperion also provides critical UV capability during the period between the end of life of the Hubble Space Telescope and the start of science for the Astro2020-recommended large IR/visible/UV flagship space telescope in the mid-2040s (Astro2020 Figure S.1). Hyperion's high-resolution, wide-field spectrograph partly closes this gap in UV access, enabling groundbreaking discoveries well beyond the mission's three science objectives.

\section*{Acknowledgments}

The Hyperion mission is a collaboration between the University of Arizona, NASA's Jet Propulsion Laboratory California Institute of Technology, and Ball Aerospace. Each organization provided seed funding for mission development, both in the 2019 SMEX version and the 2021 MIDEX version described here. The Hyperion team thanks each organization for their support in developing this mission concept. Science team members at other institutions contributed their time to the mission concept, and the Hyperion team is grateful for their work. Graphics support was provided by Eva Grall, Katie Peek, and Mark Seibert. The Hyperion team thanks them for their contributions.

This research was carried out at the Jet Propulsion Laboratory, California Institute of Technology, under a contract with the National Aeronautics and Space Administration (80NM0018D0004)

TJH is funded by a Royal Society Dorothy Hodgkin fellowship.


\bibliography{report}   
\bibliographystyle{spiejour}   

\listoffigures
\listoftables

\end{spacing}
\end{document}